\documentclass[10pt,twocolumn]{article}
\usepackage[T1]{fontenc}
\usepackage[letterpaper,margin=1in]{geometry}
\usepackage{microtype}
\usepackage{array,booktabs}
\usepackage{graphicx}
\usepackage[hidelinks]{hyperref}
\newcolumntype{L}[1]{>{\raggedright\arraybackslash}p{#1}}
\title{Testing Settlement Flags after Late Evidence and Process Restart}
\author{%
  Zengjian Liu\\Peachy Lab \and
  Tommy He\\Peachy Lab \and
  ELE Zhao\\Peachy Lab \and
  HongLiang Sun\\Peachy Lab
}
\date{}
\hypersetup{pdftitle={Testing Settlement Flags after Late Evidence and Process Restart},pdfauthor={Zengjian Liu, Tommy He, ELE Zhao, HongLiang Sun}}

\begin{document}
\maketitle
\begin{abstract}
A settlement flag can survive an application restart while the economic effects remain ambiguous. We define a contract linking the fixed plan, source admission at two cutoffs, identity-keyed ledger effects and post-invocation state. We generated 504 synthetic histories, ran each through a file implementation and a separately coded SQLite implementation, and observed 672 controlled process exits. The file implementation missed 54 same-identity payment-date conflicts, including 27 that credited cash early, and a monitor using run flags, artifact presence and purchase counts accepted all 54. The same contract distinguishes a purchase awaiting closure from a correct quote wait for a single writer on a running kernel and filesystem.
\end{abstract}

\section{Introduction}

A settlement flag records that a workflow reached a particular step. Its operator may need a different answer: whether a purchase already changed the account, whether the booked effects use evidence admissible at the relevant time, and whether a retry should place another purchase. After an application exits, a waiting flag can coexist with a saved purchase whose archival step is unfinished. The same flag can also accompany an unmet quote requirement. Late-arriving data add a second ambiguity. A settlement step can finish while a previously booked receivable has been changed by an inconsistent delivery.

Consider two reports for the same distribution. Both give the same amount. The second gives an earlier payment day. A check based on the identifier and amount may accept the later report and record cash before the day supported by the first version. An operator checking only the final flag or the number of purchases misses the difference. As a separate example, two distributions can share a record and payment date while representing distinct entitlements. Ford's public declaration of a regular and a supplemental distribution illustrates this shape. Our cases use synthetic values and specify effective dates independently of that declaration~\cite{ford}.

We ask which observable effects distinguish a valid settlement, a purchase awaiting closure, and a correct wait. The study contributes a contract, a generator and two comparisons.
\begin{samepage}
\begin{enumerate}
\item A contract that links a fixed decision and quote session to source eligibility, identity-keyed amounts, and the post-invocation run flag.
\item A bounded generator and a deterministic oracle, paired with a portable file implementation and a separately coded SQLite implementation. The observations come from fresh reader processes after each attempt.
\item An evaluation of two operational diagnostics and of verification around archive creation. It identifies a case in which the file implementation books an unsupported payment under the tested stable-identity policy.
\end{enumerate}
\end{samepage}

\section{Related work}

Snodgrass develops the distinction between valid time and transaction time in temporal databases~\cite{snodgrass}. Clifford and Isakowitz also distinguish the reference time used to interpret a database fact~\cite{clifford}. We use an effective day and a recording time for ledger projection, and acquisition and publication times for source admission. Lamport distinguishes causal order from physical clock readings~\cite{lamport}. In our single-writer histories, the runner controls the sequence of application milestones and the input timestamps determine which evidence is admissible.

Linearizability constrains concurrent observations by requiring each operation to appear to take effect between invocation and response~\cite{herlihy}. We prescribe sequential attempts and inspect the effects that remain visible across process lifetimes. RIFL binds each request identity to a completion record so retries can recover earlier results without duplicating effects~\cite{rifl}. This motivates following the purchase identity through successive observations. Gray's transaction concept gives atomicity to a group of effects~\cite{gray}, and ARIES implements transaction recovery with write-ahead logging~\cite{aries}. We check the observed group of ledger effects separately from the workflow's completion flag.

Stonebraker et al. distinguish durable execution from workflow correctness and study rollback and compensation~\cite{stonebraker}. Beldi logs function operations to support fault-tolerant and transactional workflows~\cite{beldi}. Boki builds fault-tolerant workflow support over a shared log~\cite{boki}. In keeping with the end-to-end argument of Saltzer et al., our observation contract checks the application's intended changes to cash and receivables at the point where they are used~\cite{saltzer}.

Pillai et al. show how crash consistency depends on filesystem persistence properties~\cite{pillai}. Mohan et al. find bugs by generating bounded filesystem workloads and simulating crashes~\cite{mohan}. FSCQ proves recovery properties from explicit crash conditions~\cite{fscq}. IFSCQ verifies detection of corrupted disk contents under an untrusted-storage model~\cite{ifscq}. Our file integrity controls compare application-level bytes on a running system. The restart experiment holds the filesystem and kernel live throughout.

QuickCheck makes properties and data generators executable~\cite{quickcheck}, and Csmith constructs generated programs with well-defined behavior for comparison~\cite{csmith}. CHESS controls concurrent schedules to reproduce failures~\cite{chess}. FoundationDB exercises faults in deterministic simulation~\cite{foundationdb}, and Newcombe et al. use model checking to find errors in distributed system designs~\cite{newcombe}. Lineage-driven fault injection uses dependencies to select failures that could prevent an expected result~\cite{ldfi}. We specify a finite grid, cause exits at explicit application boundaries and record the resulting effects through fresh processes.

\section{Observable contract}

\subsection{Admission and accounting}

A prepared run fixes the decision and named execution session before later evidence is processed. A source $s$ has expected digest $h_s$, raw bytes $b_s$, acquisition time $a_s$, and optional publication time $p_s$. At cutoff $t$, the source is eligible when
\[
  \begin{array}{l}
  H(b_s)=h_s\quad\land\quad a_s\mbox{ exists and }a_s\le t\\
  \quad\land\quad(p_s\mbox{ absent or }p_s\le t).
  \end{array}
\]
The two cutoffs are the original decision and the later delivery. Disagreement between a document's source identity and the enclosing report stops accounting even when the time and byte check passes.

Each ordinary distribution has an official identifier, ex-date, pay date and per-share amount. The first supported terms fix its identity in the test contract. A compatible redelivery can add a payment that has become due, with a single effect per identity and kind. A change of amount or either date is an explicit conflict and adds no effect. A new identifier represents a distinct entitlement even on the same day. For a query at effective day $d$ and knowledge time $t$, a booked effect contributes when its effective day is at most $d$ and its recording time is at most $t$. The tested deliveries occur on or after the ex-date, using holdings established before that date. Recognizing a supported distribution creates a receivable, which stays outstanding while the first supported pay date is in the future. Recording a payment when due reduces the receivable and credits cash. Deliveries before entitlement and reversals after authoritative corrections require additional rules outside the tested domain.

\subsection{Execution and closure}

The expected buy uses a valid capture for the fixed execution session and the stored plan. During an inspection, capture, execution, archive and run flag are separate observations. The expected effect is a single buy whose session, amount and quantity are all checked. When the first inspection after an exit already shows the buy, later inspections check its identity and ledger contribution without counting the repeated observations as trades. A missing usable quote leaves the run waiting with zero buys, even when an earlier distribution has already changed the account. An archived result can be blocked, so its existence is also distinct from a purchase.

\section{Generated evaluation}

Following the distinction between test selection and execution in model-based testing~\cite{utting}, we specify the history space and the process boundaries separately. The generator composes ten evidence families, up to three timestamp profiles, two delivery orders and nine invocation schedules. The families include distinct same-day identities, repeat delivery, conflicting amount or date, an inverted date, an inconsistent source identity, and missing or inconsistent evidence metadata. An exact-cutoff case fixes equality as a positive control. Late acquisition and publication vary separately. Each of the nine schedules uses four attempts with a new process on each attempt. Six schedules stop the first worker at an application boundary. We verify each stop from the process exit code and the recorded boundary. The other schedules cover temporary and terminal quote gaps and uninterrupted completion.

The file implementation saves its purchase and archive separately. The SQLite implementation uses an identity table, a per-effect primary key, a separate trade row and transactional checkpoints~\cite{sqlite}. The accounting and projection code are separately written. Barr et al.\ describe the test oracle problem as deciding whether observed behavior is correct~\cite{barr}. Our oracle derives expected postings from the first supported terms, sums visible effects in integer cents, and uses a milestone specification for each schedule. The executor receives the input only. Agreement between the two implementations provides a cross-implementation check over this contract.

Each history supplies both implementations with the same input. Fresh readers reconstruct the initial and four post-attempt observations. An exact pass requires agreement on source admission and accounting dispositions, identity-keyed events, cash, receivables, quantity and session, actual process exit, preserved plan identity and every observed phase. We compare the prespecified contract to two narrower diagnostics. A terminal-flag test treats \texttt{SETTLED} as a pass. A stronger test checks the expected flag, file or record presence and purchase count after each attempt. It has access to the prescribed schedule and recognizes correct waiting. Both diagnostics leave out source disposition, amount and identity.

The unit is the input history, paired across the two methods. A 19-history pilot informed the 504-history design. The latter excludes redundant clock variations where a field is missing. A separate AI-assisted review session approved the manifest linking the protocol, cases and programs before execution of the 1,008 formal cells. A third-party timestamp attests to the manifest digest. The first execution per cell supplies the main score, with ten prespecified histories repeated once per method in fresh directories. A separate 30-cell comparison changes saved records before verification. The appendix reports the design denominators by evidence family and invocation schedule.

\section{Results}

\subsection{Consequences of a payment date conflict}

All 1,008 cells yielded complete observations. The file implementation met the contract in 450 of 504 histories, and the SQLite implementation in 504 of 504. The 54 discordant pairs all involved the same identifier, ex-date and amount with a changed pay date. The file path deduplicates effects by identity and effect kind and checks that the amount agrees. It accepts the changed date without recording the expected conflict. When the later payment day arrives first, the second report causes all 27 such file runs to credit 400 cents of cash early and reduce receivables by 400 cents. In the reverse order, the valid earlier payment already exists. All 27 file runs in that order have the expected cash and receivables and miss the conflict disposition. The two consequences each appear across all three timing profiles and nine schedules. The book difference is visible before the first purchase attempt, including in the absence of an injected exit.

Table~\ref{tab:diagnostic} shows how the narrow monitors score against the contract. The terminal flag falsely passes 48 of the 54 file-path conflicts and falsely rejects 50 file histories and 56 SQLite histories that correctly wait. Six file-path conflicts end with a waiting flag in the terminal-gap schedule. The stage-level check predicts the intended waiting states, yet falsely passes all 54 conflicts because the flags, artifact presence and purchase counts agree with the expected milestones.

\begin{figure*}[t]
\centering
\includegraphics[width=\textwidth]{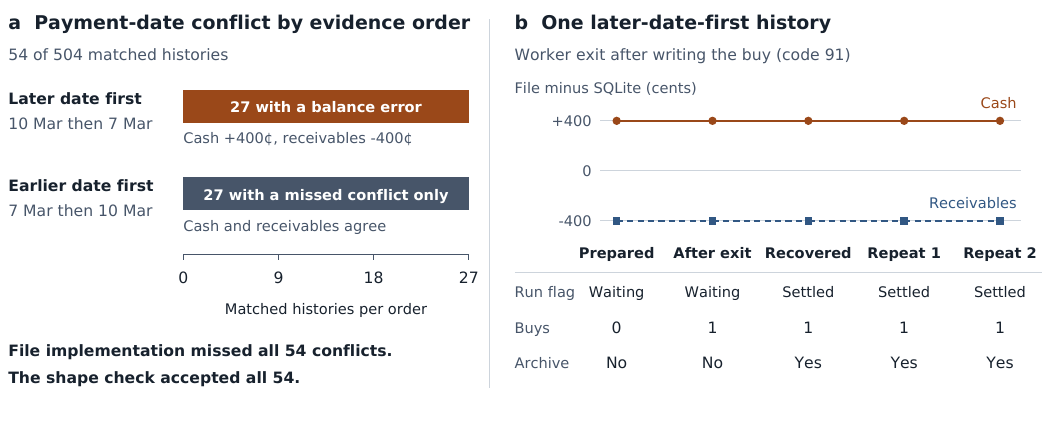}
\caption{A changed payment date in a study of 504 matched histories, each run in both implementations. (a) The 54 histories with a changed payment date split into 27 per delivery order. Balance differences are measured at the initial inspection. Conflict and shape outcomes refer to the full history. (b) Five inspections of one of the 27 later-date-first histories, selected at the exact cutoff with an exit after writing the buy. The balances are file minus SQLite. Both implementations show the displayed run flags, buy counts and archive presence. Both panels draw on the same 504 histories partitioned in Table~\ref{tab:coverage}.}
\label{fig:payment-date}
\end{figure*}

Figure~\ref{fig:payment-date}(a) shows why the two balances distinguish only half of these conflicts. In the earlier-date-first histories, the recorded amounts agree because the first report already supports the payment. The missing conflict disposition is the only observed difference in those 27 histories.

\begin{table*}[t]
\centering\small
\begin{tabular}{@{}lrrrrr@{}}
\toprule
Implementation & Exact pass & Flag: false pass & Flag: false reject & Shape: false pass & Shape: false reject \\
\midrule
JSON files & 450/504 & 48 & 50 & 54 & 0 \\
SQLite & 504/504 & 0 & 56 & 0 & 0 \\
\bottomrule
\end{tabular}
\caption{Formal conformance and diagnostic disagreements over the same 504 synthetic histories per implementation. Each false pass or false reject is relative to the full contract. Shape checks the expected stages, stored artifact presence and purchase counts.}
\label{tab:diagnostic}
\end{table*}

\subsection{Waiting after a buy}

The main runs reached all six specified exit boundaries, with 112 worker exits at each. Across both implementations, 112 first inspections after the execution write showed a buy and a waiting flag with the archive pending. Another 112 first inspections after the archive write showed a buy and an archive with the same waiting flag. With a terminal quote gap, 112 cells showed waiting without a buy or an archive. Later attempts preserved the published purchase when closing the interrupted runs. Across all schedules, no inspection contained more than one buy. The normalized observations matched the first execution in all 20 prespecified repeat cells.

In Figure~\ref{fig:payment-date}(b), the balance gap is already present before either implementation attempts the buy. After the exit, both readers see one buy while the run flag still reads waiting and the archive is absent. The next inspection shows both runs settled and archived. The file implementation still shows 400 cents too much cash and 400 cents too little in receivables, and the gap persists at both subsequent inspections. In this history, recovery closes the purchase workflow while the earlier accounting error remains visible.

\subsection{Archive-dependent verification}

All 30 integrity controls yielded observations. With the archive pending, recomputation refused all five material changes. Reuse refused the two changes that failed parsing or plan binding and retained the execution under a changed quote, changed buy amount or missing capture. With an archive present, both policies refused all six applicable material changes and a formatting-only change that altered the digest. The byte check can therefore block a semantically equivalent file. The settled flag from the earlier attempt remained after each refusal with an existing archive. Both policies accepted the intact controls and share the same accounting implementation.

\begin{table}[t]
\centering\small
\begin{tabular}{@{}lrr@{}}
\toprule
Archive at start & Recompute & Reuse \\
\midrule
Pending & 5/5 & 2/5 \\
Present & 6/6 & 6/6 \\
\bottomrule
\end{tabular}
\caption{Material alterations refused in each synthetic integrity context. There are 30 total method cells including the unchanged and formatting-only controls. Denominators count applicable material changes within a context and policy.}
\label{tab:integrity}
\end{table}

\subsection{A smaller comparator}

A separately developed comparator also checks portability of the accounting definitions. A Claude model received only the synthetic contract and its input format. We then ran the returned Python program locally on all 65 histories. A separately written set-based scorer checked the outputs. The program matched every expected source decision, projection and stipulated phase observation. This companion contract covers accruals, missing documents, digest mismatches, duplicates and conflict quarantine. Its restart observations are logical, and its conflict rule excludes an ambiguous identity. The comparison supports implementability for this smaller contract. The process-persistence and payment-date results come from the two implementations in the main experiment.

In the earlier synthetic study (Appendix~\ref{sec:historical}), recomputation used 70 calculations versus 26 for reuse, with 41 of the 44 extra calls after the first settled inspection.

\section{Interpretation and limitations}

The changed-date finding shows that a usable quote, a single purchase and a settled flag can coexist with a prematurely discharged receivable. A verifier needs to check the recorded effects against the first supported terms. A workflow waiting with a purchase already recorded calls for closure, while a workflow waiting for a quote still lacks the prerequisite for execution. The integrity comparison identifies a third condition in which an existing archive supplies digests for verification.

The first-supported-terms rule is part of this synthetic contract. An application that permits authoritative corrections would need explicit rules for amendments, reversals and reconciliation. The finite grid provides evidence about declared mechanisms and conditions. It does not estimate how often an error occurs with live inputs. We exercise a single writer with the kernel and filesystem still running, after exits at observed application boundaries. The results do not cover a power cut, interrupted storage writes, concurrency or native replay of the historical private implementation. The same research team developed both implementations and the oracle, so all three could share a contract error even though the implementations account for events differently.

\section{Use of AI}

AI tools assisted coding, synthetic case preparation, analysis, and manuscript drafting. Earlier expected outcomes were prepared in a separate session using the same model family as the work session. A Claude model, identified by the service as \texttt{claude-sonnet-5-5-low}, generated the separate comparator from a synthetic contract. The comparison runs and scoring used deterministic code without model inference.

\section{Artifact availability}

The authors retain the runnable implementations, generator, deterministic oracle, prespecified design, and normalized observations. The retained instructions cover rebuilding the design, rescoring the archived observations, and replaying the formal experiment and pilot in fresh directories. The experiments use synthetic sources and standard-library Python with SQLite.

The SHA-256 manifest identifies the plan, code and cases. The pre-run approval names the digest, and the third-party timestamp attests to it at a specified time. The historical table retains the original case identifiers. The complete historical scorer and private implementation remain in a separately controlled archive.

\clearpage
\onecolumn
\appendix
\section{Formal coverage}

Table~\ref{tab:coverage} reports the same 504 paired histories by evidence family and by invocation schedule. Each group of six file-path mismatches in a schedule contains three of each delivery order. The 20 repeat cells and 30 integrity controls have separate denominators.

\begin{table}[!ht]
\centering\small
\begin{minipage}[t]{0.48\textwidth}
\centering
\begin{tabular}{@{}lrr@{}}
\toprule
Evidence family & JSON files & SQLite \\
\midrule
Distinct same-day IDs & 54/54 & 54/54 \\
Different effective dates & 54/54 & 54/54 \\
Exact redelivery & 54/54 & 54/54 \\
Conflicting amount & 54/54 & 54/54 \\
Changed payment date & 0/54 & 54/54 \\
Payment before entitlement & 54/54 & 54/54 \\
Inconsistent source identity & 54/54 & 54/54 \\
Optional publication time & 36/36 & 36/36 \\
Missing acquisition time & 36/36 & 36/36 \\
Digest mismatch & 54/54 & 54/54 \\
\bottomrule
\end{tabular}
\end{minipage}\hfill
\begin{minipage}[t]{0.48\textwidth}
\centering
\begin{tabular}{@{}lrr@{}}
\toprule
Invocation schedule & JSON files & SQLite \\
\midrule
Uninterrupted & 50/56 & 56/56 \\
Temporary quote gap & 50/56 & 56/56 \\
Terminal quote gap & 50/56 & 56/56 \\
Exit before capture & 50/56 & 56/56 \\
Exit after stale capture & 50/56 & 56/56 \\
Exit before execution & 50/56 & 56/56 \\
Exit after execution & 50/56 & 56/56 \\
Exit after archive & 50/56 & 56/56 \\
Exit after logging failure & 50/56 & 56/56 \\
\bottomrule
\end{tabular}
\end{minipage}
\caption{Passes under the full contract over planned cells. Each panel partitions the same histories along a different design axis.}
\label{tab:coverage}
\end{table}

\section{Historical observations}
\label{sec:historical}

An earlier set of 18 synthetic histories produced 54 method cells. Both policies in the same reference implementation passed the earlier contract on all 18 histories. A separately implemented private path passed 16. These cases motivated the current study. The test rules of the current generated study are stated separately.

\begin{table}[!ht]
\centering
\begin{tabular}{@{}llL{0.62\textwidth}@{}}
\toprule
Case & Recorded state & Observed consequence \\
\midrule
H07 & Settled & The private implementation bought the planned quantity and retained 5.60 of 14.40 in expected receivables, omitting the distinct same-day amount of 8.80. \\
H09 & Partial & The private path showed a blocked execution and an archive, with zero buys. Cash of 9.00 was credited early and a separate receivable of 3.00 was omitted. \\
H18 & Waiting & All three paths observed repeated missing or stale quotes. The original contract expected waiting with zero buys. \\
\bottomrule
\end{tabular}
\caption{Three historical cases. Amounts are synthetic dollars. The authors retain their derived observations. The private executable remains restricted.}
\end{table}

The earlier expected results were authored in a separate session using the same model family as the work session. The output was evaluated by deterministic code against an answer committed before execution in local records. These records establish the order within the local procedure. They do not provide an external registration timestamp. No model runs inside the execution or scoring harness.

\end{document}